%Paper: 9108020
%From: Jeff Harvey <harvey@curie.uchicago.edu>
%Date: Mon, 26 Aug 91 11:18:53 CDT

\input phyzzx
\input jnl
\input reforder
\ignoreuncited
EFI-91-30

\title            STRING THEORY AND THE DONALDSON POLYNOMIAL

\author           Jeffrey A. Harvey%PUT THE AUTHOR'S NAME ON THIS LINE

\affil            Enrico Fermi Institute, University of Chicago
                  5640 Ellis Avenue, Chicago, IL 60637
                  Internet: harvey@curie.uchicago.edu

\author           Andrew Strominger

\affil            Department of Physics, University of California
                  Santa Barbara, CA 93106
                  Bitnet: andy@voodoo

\abstract         It is shown that the scattering of
spacetime axions with fivebrane solitons of heterotic
string theory at zero momentum is proportional to the Donaldson polynomial.

\vskip 1.5in
\noindent July, 1991

\endtitlepage

\head{\bf I. Introduction}  %YOU CAN REPLACE I. Introduction WITH ANY HEADING.
                        %THE { AND } ARE ESSENTIAL.

A ${p}$-brane (i.e. an extended object with a ${p} + {1}$-dimensional
worldvolume) naturally acts as a source of a ${p} + {2}$ form
field strength ${F}$ via the relation
$$
{\nabla^M}{F}_{MN_1}{\cdots}_{N_{p+1}} = {q}{\Delta}_{{N_1}{\cdots}
{N_{p+1}}}\eqno(div)
$$
\noindent where ${\Delta}$ is the ${p}$-brane volume-form times a
transverse ${\delta}$-function on the ${p}$-brane. In $d$ dimensions they
can therefore carry a charge
$$
{q} = {\int}_{\Sigma^{d-p-2}} *{F}.\eqno(qcharge)
$$
where the integral is over a $d-p-2$ dimensional hypersurface at spatial
infinity.
\noindent The dual charge
$$
{g} = {\int}_{\Sigma^{p+2}} {F}\eqno(gcharge)
$$
\noindent can be carried by a ${d-p-4}$ brane. A straightforward generalization
[\cite{teinep}] of Dirac's original argument implies that quantum mechanically
the
charges must obey a quantization condition of the form
$$
{qg} = {  n},\eqno(quant)
$$
\noindent just as for the special case of electric and magnetic charges
in ${d} = {4}$. In particular, strings in ten dimensions are
dual, in the Dirac sense, to fivebranes. Thus fivebranes are the magnetic
monopoles of string theory.

In [\cite{STR.55,strb}] it was shown that heterotic
string theory admits exact fivebrane soliton solutions.
The core of the
fivebrane consists of an ordinary Yang-Mills instanton. Thus heterotic
strings are dual, in the Dirac sense, to Yang-Mills instantons.

This simple connection between Yang-Mills instantons and
heterotic string theory raises many possibilities.  On the one hand, heterotic
string theory might be used as a tool to study the rich mathematical
structure of Yang-Mills instantons, or to suggest interesting generalizations.
On the other hand, the mathematical results of Donaldson [\cite{donref}]
and others on
the construction of new smooth invariants for four-manifolds
may have direct implications for non-perturbative semi-classical
heterotic string theory.

In this paper this connection is elucidated as follows. We consider
${N}$ parallel fivebranes on the manifold ${M^6}{\times}{X}$ where
${M^6}$ is six-dimensional Minkowski space and ${X}$ is the
four-manifold transverse to the fivebranes. (The consistency of string
theory requires ${c_1} ({X}) {\geq} {0}$, though this condition might
conceivably be relaxed in the present context by allowing singularities
along the divisor of ${c_1}$.)
The quantum ground states of this system are found
to be cohomology classes on the ${N}$-instanton moduli space ${\cal M}_{N}
({X})$. Transitions among these ground states may be induced by
scattering with a zero-momentum axion. Such axions are characterized
by a harmonic two-form or an element of
${H^2}({X})$, and the ${S}$-matrix then defines
a map ${H^2}({X}){\rightarrow}{H^2}({\cal M}_{N})$.
This map turns out to be precisely
the Donaldson map.
The fact that the scattering
is a map between cohomology classes is ultimately a consequence of
zero-momentum spacetime supersymmetry.
Multiple axion scattering is given by the
intersection numbers on ${\cal M}_{N}$ of these elements of
${H^2}({\cal M}_{N})$, which is the Donaldson polynomial.

This representation of the Donaldson map as a string $S$-matrix element
leads to an apparently new geometrical interpretation of Donaldson theory.
For any K\"ahler manifold $X$
%with K\"ahler form $J={i \over 2}
%\partial \bar \partial K$,
there is a K\"ahler geometry on
${H}^2(X,C)\times {\cal M}_N(X)$ with K\"ahler potential
defined by
$$
{\cal K}={1 \over 2}\int_X E \wedge J-{\rm ln}\int_X J\wedge J
\eqno(why)
$$
where $J$ is the K\"ahler form on $X$ and $E$ is a solution of
$$
{\rm tr}F \wedge F=i \partial \bar \partial E.
\eqno(ffe)
$$
$E$ is an analog of the Chern-Simons form for K\"ahler geometry.
The Donaldson map is then given by a mixed component of the
Christoffel connection, computed as the third derivative of
$\cal K$.

It is noteworthy that the final expressions we derive for the Donaldson
map and polynomial are similar
to those given by
Witten [\cite{WIT.2}].
Indeed, the embedding of four-dimensional Yang-Mills instantons into
ten-dimensional string theory given by fivebrane solitons seems
to produce a structure of zero-momentum fields and symmetries similar,
if not identical, to that of Witten's topological Yang-Mills theory.
Though we have not done so, it is possible that the complete structure
of topological Yang-Mills theory can be derived from zero-momentum
string theory in the soliton sector. This is perhaps in contrast to the usual
notion [\cite{WIT.2}] that topological field theory is relevant to a
short-distance phase of string theory.

We work only to leading order in $\alpha^{\prime}$
in this paper. An interesting question,
which we do not address, is whether or not higher-order or non-perturbative
corrections provide a natural deformation of the Donaldson polynomial
analogous to the deformation of the cohomology ring provided by string theory.

This paper is organized as follows. In Section II we establish our
notation and review some properties of instanton moduli space.
The collective coordinate expansion leading to the low-energy effective
action is derived in some detail, and is then used to characterize
the N-fivebrane ground states. In Section III.A the collective coordinate
expansion is continued to reveal the Donaldson map as a subleading term in
the low-energy effective action.
Section III.B gives an alternate derivation of the Donaldson map using
supersymmetry and K\"ahler geometry, and derives (\call{why}). In
Section III.C we discuss the representation of the Donaldson map
as a period of the second Chern class
which may be relevant in the present context. In Section III.D
the string scattering amplitude which measures the Donaldson
polynomial is described. We conclude with discussion in Section
IV.

\head{\bf II. INSTANTON MODULI SPACE AND THE COLLECTIVE
COORDINATE EXPANSION}
The derivation of the Donaldson map and polynomial from ten-dimensional
string theory is straightforward though somewhat involved. We
begin with the action describing the low-energy limit of heterotic string
theory:
$$\eqalignno{
{S_{10}} = {{1}\over{{\alpha^{\prime}}\,^{4}}} &{\int} {d^{10}}{x}  {\sqrt{-g}}
\,\,{e^{-2\phi}} \bigl( {R} + {4}{\nabla_M}{\phi}{\nabla^M}{\phi}
{-}{{1}\over{3}}{H_{MNP}} {H^{MNP}} -
{\alpha^{\prime}}{ \rm tr}{F_{MN}}{F^{MN}}\cr
&{-}{\bar\psi}_M\,{\Gamma^{MNP}}{\nabla_N}{\psi_P} +2 {\bar\lambda}
{\Gamma^{MN}}{\nabla_M}{\psi_N} +  {\bar\lambda}\,{\Gamma^M}
{\nabla_M}{\lambda}
-2{{\alpha^{\prime}}}\,{\rm tr}{\bar\chi}\,{\Gamma^M}{D_M}{\chi} \cr & -
{\alpha^{\prime}}
{\rm tr}\,{\bar\chi}\,{\Gamma^M}
{\Gamma^{NP}} {F_{NP}}
({\psi_M} + {{1}\over{6}} {\Gamma_M}{\lambda})
{+}2 {\bar\psi}_N\,{\Gamma^M}{\Gamma^N}{\lambda}{\nabla_M}{\phi} -2
{\bar\psi}_M {\Gamma^{M}}{\psi^N}{\nabla_N}
{\phi}\cr
&{+}{{1}\over{12}}
{H_{MNP}} ({2\alpha^{\prime}}{\rm tr}{\bar\chi}{\Gamma^{MNP}}{\chi} -
{\bar\lambda}
{\Gamma^{MNP}}{\lambda} + {\bar\psi}^S\,{\Gamma_{[S}}
{\Gamma}^{MNP}\,{\Gamma}_{T]}{\psi^T}\cr
& \qquad \qquad \qquad +2 {\bar\psi}_S\,{\Gamma^{SMNP}}{\lambda})+{\cdots}
\bigr)&(sten) \cr}
$$
\noindent where ``${+}{\cdots}$'' indicates four-fermi as well as higher-order
${\alpha^{\prime}}$ corrections, ${H} = {dB} + \alpha^{\prime}
{\omega_{3L}} - \alpha^\prime
{\omega_{3Y}}$, $\omega_{3L}$ and $\omega_{3Y}$ are the Lorentz and
Yang-Mills Chern Simons three-forms respectively,
and ``tr'' is ${{1}\over{30}}$ the trace in the adjoint representation
 of ${E_8}\times {E_8}$ or SO(32).
A supersymmetric solution of the equations of motion following
from (\call{sten}) is one for which there exists at
least one Majorana-Weyl spinor ${\eta}$ obeying
$$\eqalignno{
\delta \psi_M &= {\nabla_M}{\eta} - {{1}\over{4}}{H_{MNP}}\Gamma^{NP}
{\eta}={0}\cr
\delta \lambda &=
{{1}\over{6}}{H_{MNP}}
{\Gamma^{MNP}}{\eta} -
{\nabla_M}{\phi}{\Gamma^M}{\eta}
={0}\cr
\delta \chi &= -{{1}\over{4}}
{F_{MN}}{\Gamma^{MN}}{\eta}={0.}&(susy)\cr}
$$
\noindent The general solution of this form on ${X}{\times}{M^6}$,
where ${X}$ is a K\"ahler manifold with ${c_1} {\geq} {0}$ and ${M^6}$ is
flat six dimensional Minkowski space,
was found in [\cite{STR.26}]. The gauge
field may be any self-dual connection on ${X}$:
$$
{F_{\mu\nu}} = {{1}\over{2}}{\epsilon_{\mu\nu}}\,^{\rho\sigma}\,
{F_{\rho\sigma}}\eqno(sdual)
$$
\noindent where ${\mu},{\nu}$ are indices tangent to ${X}$.  Let
${\hat g}$ be a Ricci flat K\"ahler metric on ${X}$.
Then the dilaton is
the solution of
$$
{\hat{\square}}\,{e^{2\phi}} = {\alpha^{\prime}} ({\rm tr}\,{\hat R}_{\mu\nu}
{\hat R}^{\mu\nu} - {\rm tr}\,{F_{\mu\nu}}\,{F^{\mu\nu}})\eqno(deom)
$$
\noindent and the metric and axion are
$$\eqalignno{
{g_{\mu\nu}}=&{e^{2\phi}}\,{\hat g}_{\mu\nu}\cr
{H_{\mu\nu\lambda}}=& - {\epsilon_{\mu\nu\lambda}}^{\rho}{\nabla_\rho}
{\phi}\cr
{g_{ab}}=&{\eta_{ab}}&(metax)\cr}
$$
where ${a},{b}$ are tangent to ${M^6}$.
Special cases of this general
solution are discussed in [\cite{STR.55,sjrey,strb}].
In [\cite{wbrane}] it was argued for $X=R^4$ that such solutions are in one to
one
correspondence with
exact solutions of heterotic string theory.
For ${c_2}({F}) = {N}$, this may be viewed as a configuration of
${N}$ fivebranes transverse to ${X}$.
(It may also be viewed as a ``compactification'' to six dimensions,
though ${X}$ need not be compact.)
Since (given the metric on ${X}$)
there is one unique solution for every self-dual Yang-Mills connection
(\call{sdual}), the space of static ${N}$ fivebrane solutions is identical
to the moduli space ${\cal M}_{N}$ of ${N}$-instanton configurations
on ${X}$.

For $c_1(X) \ge 0$ and $c_2(R) \ge c_2(F)$, the
metrics $g$ of (\call{metax}) are
geodesicaly complete, but may be non-compact.
If  ${c_1}(X)>0$, there
are geodesically complete but non-compact Ricci-flat
metrics with bounded curvature [\cite{yau}].
This may be viewed as a singular metric on
$X$ or a non-singular metric on $X$ minus the divisor of $c_1$.
There appears to be no special difficulty in defining string propagation on
such geometries (though they would not be suitable for Kaluza-Klein
compactification). Singularities may also arise in solving
(\call{deom}). If $c_2(R)>c_2(F)$, the singularities are of the type studied
in [\cite{strb}] and again produce no difficulties.

On the other hand if $c_1<0$ or $c_2(R)<c_2(F)$, the metric in (\call{metax})
may have real curvature singularities, which could potentially render
string theory ill-defined. More work must be done before our methods can be
used to directly study these cases, but the validity of our final formulae
for all K\"ahler $X$ suggests that it may be possible to do so. Possible
approaches would be to consider the more general case of time-dependent
metrics, or to consider the (eight-dimensional) cotangent bundle of $X$
which has $c_1=0$.

The solutions of (\call{sdual}),(\call{deom})  have bosonic zero modes tangent
to
${\cal M}_{N}$. To leading order in ${\alpha^{\prime}}$ these zero modes
involve only the gauge field and will be denoted ${\delta_i}{A_\mu}({x})$,
where ${i} = {1,}{...,}{m},\,\,{m}\,\,{\equiv}\,\,
{\rm dim} ({\cal M}_N)$ and $x$ is a coordinate on
$X$. The zero modes obey the linearized self-duality equation
$$
{D}_{[\mu}{\delta_i}{A_{\nu]}} = {{1}\over{2}}{\epsilon_{\mu\nu}}
\,^{\rho\sigma}\,{D}_{\rho}{\delta_i}{A_{\sigma}}.\eqno(aeom)
$$
%The number of independent solutions of (\call{aeom}) modulo gauge
%transformations can be determined using the Atiyah-Singer index theorem
%[\cite{ahs}]. For gauge group $SU(2)$ the result is
%$$
%{\rm dim} {\cal M}_N = 8 N - {3 \over 2} (\chi(X) + \sigma(X)) \eqno(dimm)
%$$
%where $\chi$ and $\sigma$ are the Euler characteristic and signature of $X$.
For gauge groups larger than $SU(2)$ or for metrics on $X$ which are not
``generic'' the gauge connection will in general
be reducible
(there exist non-trivial solutions of $D \phi=0$).
This leads to orbifold singularities in
${\cal M}_N$. In what follows we will restrict ourselves to irreducible
connections and ignore such subtleties.

If ${Z^i}$ is a coordinate on ${\cal M}_{N}$, and
${A}^0_{\mu}({x},{Z})$ a family of self-dual connections, the zero
modes are given by
$$
{\delta}_i {A_\mu} = {\partial_i}{A^0_{\mu}} - {D}_{\mu}{\epsilon_i}
\eqno(zmode)
$$
\noindent where ${\epsilon_i}({x},{Z})$ are arbitrary gauge parameters and
${\partial_i} = {{\partial}\over{{\partial}{Z^i}}}$. It is convenient to
fix ${\epsilon_i}$ by requiring
$$
{D}^{\mu}{\delta_i}{A_\mu} = {0}\eqno(choice)
$$
so that the zero modes are orthogonal to fluctuations of the gauge
field obtained by gauge transformations.
The gauge parameter
${\epsilon_i}$ then defines a natural gauge connection on ${\cal M}_{N}$ with
covariant derivative
$$
{s_i} = {\partial_i} + [{\epsilon_i},{~~}]\eqno(sconn)
$$
\noindent which has the property
$$
[{s_i}, {D}_{\mu}] = {\delta_i}{A_{\mu}}.\eqno(sd)
$$
\noindent The Jacobi identity for ${s_i},{D}_{\mu}$ and ${D}
_{\nu}$ implies
$$
{s_i}{F}_{\mu\nu} = {2}{D}_{[\mu}{\delta_i}{A_{\nu]}}.\eqno(sdd)
$$
\noindent The Jacobi identity for ${s_i}, {s_j}$ and ${D}_{\mu}$
implies
$$
{D}_{\mu}{\phi_{ij}} = {-} {2} {s_{[i}}{\delta}_{j]}{A_{\mu}}\eqno(ssd)
$$
\noindent where
$$
{\phi_{ij}} = [{s_i},{s_j}]\eqno(phi)
$$
\noindent is the curvature associated to ${s_i}$. These relations will
be useful shortly.

A natural metric ${\cal G}$ on ${\cal M}_{N}$ is induced from the
metric $g$ on ${X}$:
$$
{\cal G}_{ij} = {\int_X}\,{d^4}{x}{\sqrt{g}}\,e^{-2\phi}{g^{\mu\nu}}{ \rm
tr}({\delta_i}
{A_{\mu}}{\delta_j}{A_\nu}).\eqno(modmet)
$$
\noindent In addition there is a complex structure $\cal J$ on ${\cal M}_{N}$
induced from the complex structure ${J}$ on ${X}$:
$$
{\cal J}_i\,^{j} = {\int_X}{d^4}{x}{\sqrt{g}}\,e^{-2\phi}{J_\mu}
\,^{\nu}\,{\rm tr}({\delta_i}{A_\lambda}
{\delta_k}{A_\nu}){g^{{\mu}{\lambda}}}{\cal G}^{kj}.\eqno(modj)
$$
\noindent It is easily seen that
the zero modes are related by
$$
{\cal J}_i\,^j\,{\delta_j}{A_\mu} = - {J_\mu}\,^\nu\,{\delta_i}
{A_\nu}.\eqno(aa)
$$
In addition to bosonic zero modes, there are fermionic
zero modes of the superpartner ${\chi}$ of ${A_{M}}$. These
zero modes are paired with the bosonic zero modes by the
unbroken supersymmetry [\cite{zum}] and are given by
$$
{\chi_{i}} = {\delta_i}{A_\mu}{\Gamma^\mu}{\epsilon^{\prime}}
\eqno(fmode)
$$
\noindent where ${\epsilon^{\prime}}$ is the
four-dimensional chiral spinor obeying
$$\eqalignno{
{J_{\mu\nu}}&= -{i}\,{\epsilon^{\prime \dagger}}{\Gamma_{\mu\nu}}\,
{\epsilon^{\prime}},\cr
{J^\mu}_{\nu}{\Gamma^{\nu}}
{\epsilon^{\prime}}&= i
{\Gamma^{\mu}}{\epsilon^{\prime}},\cr
\epsilon^\dagger \epsilon&=1.&(ee)\cr}
$$
It is easy to check, using (\call{choice}) and (\call{aeom}),
that $\Gamma^\mu D_\mu
{\chi_{i}} = {0}$.

Equation (\call{fmode}) would
appear to give ${m}$ zero modes,
where ${m}$ is the dimension of ${\cal M}_{N}$,
but we know from the index theorem that these
are not linearly independent. Using (\call{ee})
and (\call{aa}) one finds
$$
{\cal J}_i^{~j}{\chi_{j}} = {i}{\chi_{i}}.
\eqno(xx)
$$
This gives ${{m}\over{2}}$
independent zero modes, as implied by the index theorem.

The low-energy dynamics of ${N}$ fivebranes in ${X}{\times}{M^6}$ is best
described by an effective action ${S_{\rm eff}}$. This action can be
derived by a (super) collective coordinate expansion which begins
$$\eqalignno{
{A_\mu} ({x},{\sigma})&= {A^0_\mu} ({x}, {Z}({\sigma}))
+ {\cdots}\cr
{\chi} ({x},{\sigma})&= {\lambda^{i}} ({\sigma}) {\chi_{i}}
({x}, {Z}({\sigma})) + {\cdots} & (exp)\cr}
$$
\noindent where $({x},{\sigma})$ is a coordinate on ${X}{\times}{M^6}$
and the bosonic (fermionic) collective coordinates
${Z^i}\,\,({\lambda^{i}})$ are dynamical fields on the soliton
worldbrane. ${\lambda^{i}}=\lambda^i_++\lambda^i_-$ is a
doublet of six-dimensional Weyl
fermions obeying $\lambda^j=i{{\cal J}_k}^j\lambda^k$.  Under
${SO(5,1)}$ worldbrane Lorentz transformations, the
${\lambda^i}$'s transform into one another.
It is possible to assemble the
${\lambda^i}$'s into ${{m}\over{2}}$ six-dimensional symplectic
Majorana-Weyl spinors transforming covariantly under ${SO(5,1)}$. However
SO(5,1) covariance is not necessary for our purposes, and the supersymmetric
${SO(5,1)}$ covariant formulation introduces a number of extraneous
complications which obscure the connection with Donaldson theory.
Our expressions will have manifest invariance under two-dimensional
super-Poincare transformations which we take to act in the
$\sigma^0,~\sigma^1$ plane. The subscripts on $\lambda^i_\pm$
denote the corresponding two-dimensional chirality.
%While the expansion (\call{exp}) is not ${SO(5,1)}$ covariant, it is
%covariant under an ${SO(1,1)}$ subgroup which we take to act in the
%(${\sigma^{0}}, {\sigma^1})$ plane. The real and imaginary parts of
%${\lambda^i}$ are ${SO(1,1)}$ Majorana-Weyl spinors, which is summarized by
%the formula
%$$
%{i}{\Gamma^{0}}{\Gamma^1}{\lambda^i} = ({\lambda^i})^*.\eqno(twoweyl)
%$$
In accord with this and as a further simplification, we henceforth restrict
$\lambda^i$ and $Z^i$ to depend only on $\sigma^0$ and $\sigma^1$.

The effective action $S_{\rm eff}$ can be expanded in powers of inverse
length. Taking $Z^i$ to be dimensionless and $\lambda^i$ to have
dimensions of (length)$^{{-}{1 \over 2}}$ this is an expansion in the
parameter $n=n_{\partial} + n_f/2$ with $n_{\partial}$ the number of
$\sigma$ derivatives and $n_f$ the number of fermion fields. The expansion
(\call{exp}) solves the spacetime equations of motion to order $n=0$, while
the leading terms in $S_{\rm eff}$ are $n=2$. To have a consistent
action we must still solve the spacetime equations to order $n=1$.
This requires that
the component of the gauge field tangent to the worldbrane
acquires the term
$$
{A_a} ={\nabla_a}{Z^i}
{\epsilon_i}- {1 \over 2}{\phi_{ij}}
{\bar\lambda}^{i}{\Gamma_a}{\lambda}^{j}  \eqno(atan)
$$
\noindent with ${\phi_{ij}}$ given by (\call{phi}).

%Equation (\call{atan}) can also be understood as a consequence
%of low-energy supersymmetry.
%There is an ambiguity in the continuation of the expansion (\call{exp})
%corresponding to (super) diffeomorphisms of (super) ${\cal M}_N$.
%However supersymmetry singles out a preferred coordinate system: we
%demand that the worldbrane action and supersymmetry transformations take
%the canonical linear form.

The leading order worldbrane action may now be derived by substitution
of the expansion (\call{exp}) (\call{atan})
of ${A}$ and ${\chi}$ into the ten dimensional action
(\call{sten}) and integration over ${X}$, the transverse
space. Using
$$
{F_{a\mu }} = {\nabla_a}{Z^i}{\delta_i}{A_{\mu}}
-s_{[i}\delta_{j]}A_\mu\bar \lambda^i \Gamma_a \lambda^j \eqno(fza)
$$
\noindent one has the bosonic term
$$\eqalignno{
S^b_{\rm eff}&=-{2 \over \alpha^{\prime 3}}\int d^4x\sqrt{g} e^{-2\phi}
\int d^6 \sigma
\,{\rm tr}\,\,
(\delta_i A_\mu
\delta_j A_{\nu})\,\,g^{\mu\nu} \nabla_a Z^i \nabla^a Z^j\cr
&=-{2 \over \alpha^{\prime 3}}\int d^6\sigma {\cal G}_{ij} \nabla^a
Z^i \nabla_a Z^j.&(sbwb)\cr}
$$
\noindent Including the fermionic terms gives the ${d}={6}$
supersymmetric sigma model with target space ${\cal M}_{N}$:
$$
{S_{\rm eff}} = -{2 \over \alpha^{\prime 3}}
{\int}{d^6}{\sigma}{\cal G}_{ij} \bigl({\nabla^a}{Z^i}{\nabla_a}
{Z^j} + 2{\bar{\lambda}}^{i}{\Gamma^a}{(\nabla_a\lambda^j
+\nabla_a Z^k \Gamma_{kl}^j}
{\lambda^l})\bigr)
+ ({\rm fermi})^4 . \eqno(swb)
$$
Because we have maintained only an ${SO(1,1)}$
subgroup of ${SO(5,1)}$ as a manifest
symmetry of (\call{swb}), only two of the eight supersymmetries are manifest.
%These are
%$$\eqalignno{
%{\delta}{Z^i}&={\bar\epsilon}{\lambda^i}??\cr
%{\delta}{\lambda^i}&={\Gamma^a}{\nabla_a}{Z^i}{\epsilon}??&(susysix)\cr}
%$$

For values of ${Z^i}$ corresponding to ${N}$ widely separated
instantons, (\call{swb}) is approximated by ${N}$ separate terms describing
the dynamics of each of the ${N}$ fivebranes. The full action (\call{swb})
includes additional fivebrane interaction terms.

Classically, there is one static ground state for each point
${Z^i}{\epsilon}{\cal M}_{N}$. However quantum mechanical groundstates
involve a superposition over ${Z^i}$ eigenstates.
%$$
%{|}{0}\rangle  = {\int}{d^n}{Z}{|}{Z}>.\eqno(zsup)
%$$
As explained by Witten [\cite{wittena}] the supercharges of the
supersymmetric sigma model (\call{swb}) act at zero momentum as the exterior
derivative
on the target space ${\cal M}_N$, and the general
supersymmetric ground state
%The state  ${|}{0}>$ is annihilated by the supercharge ${Q}$.
%(\call{zsup}) is not the only groundstate.
%As explained by Witten [~~~~]only, the general
%groundstate
can be written in the form
$$\eqalignno{
{|}{\cal O}^s \rangle &={\cal O}^s{|}{0}\rangle \cr
{\cal O}^s&=
{\cal O}^s_{i_1}{\cdots}{}_{i_p}({Z}){\psi}^{i_1}{\cdots}{\psi}^{i_p}
&(gstate)\cr}
$$
\noindent where ${\cal O}^s_{i_1}{\cdots}_{i_p}$ is a harmonic form on
${\cal M}_{N}$, $\psi^i=Re\lambda^i_++iRe\lambda^i_-$ and
the state $|0 \rangle $ is chosen so that
$$
(\psi^i)^*|0 \rangle =0 . \eqno(vacuum)
$$

In summary, the low-energy dynamics of ${N}$-fivebranes  is
described by a supersymmetric sigma-model with target space
${\cal M}_N$ and the ground states of this system are cohomology
classes on ${\cal M}_{N}$.

\head{\bf III.THE DONALDSON MAP AND POLYNOMIAL}
In addition to the leading term (\call{swb}) in ${S_{\rm eff}}$, there are a
number
of terms representing interactions between spacetime fields which are
not localized on the fivebrane  and the localized worldbrane fields
appearing in $S_{\rm eff}$.
This corresponds to the fact that the state of the fivebrane can be
perturbed by scattering with spacetime fields. For the special limiting
case of zero-momentum spacetime fields, energy conservation implies
that scattering can only induce transitions among the groundstates.
Zero momentum spacetime axions are characterized by harmonic forms on
${X}$, so axion scattering is a map involving ${H}({X})$ and
${H}({\cal M}_N({X}))$. This strongly suggests that the
scattering should be given by the Donaldson map. In the following two
subsections we demonstrate that this is indeed the case by two separate
methods. Subsection (A) contains a straightforward continuation of the
collective coordinate expansion. In subsection (B), it is observed that
the Donaldson map has a geometric interpretation as a certain connection
coefficient derivable from a K\"ahler potential. It's form is then deduced in
a few lines using supersymmetry.

\item{\bf A.} {\bf Derivation by Collective Coordinate Expansion}

Consider the interaction of a low-momentum
spacetime axion with the worldbrane fermions.  Other
spacetime fields can be treated in an analogous fashion.
Spacetime axions are described by the potential
$$
{B_{\mu\nu}=Y(\sigma)T_{\mu\nu}}\eqno(axion)
$$
\noindent where ${T}$ is a harmonic two form on ${X}$, and $Y$
depends only on ${\sigma^0}, {\sigma^1}$.
The ten-dimensional coupling
$$
{\cal L}_{\rm int}^\prime = {1 \over{{2}{\alpha^{{\prime}{3}}}}}e^{-2 \phi}
{\partial_M}{B_{NP}}\,{\rm tr}{\bar\chi}{\Gamma^{MNP}}{\chi}\eqno(hxx)
$$
\noindent appearing in (\call{sten}) descends to a coupling in
${S_{\rm eff}}$ between one spacetime axion and two worldbrane fermions.
Substituting (\call{axion}) and (\call{exp}) into (\call{hxx}) and
integrating out the zero mode wave function one finds
$$\eqalignno{
{\cal L}_{\rm int}^\prime= & {{1}\over{{2}{\alpha^{{\prime}{3}}}}}
{\int}{d^4}{x}{\sqrt{g}}e^{-2\phi}{T_{\mu\nu}}\,{\rm tr}\,
({\bar\lambda}^{i}
{\chi}^{\dag}_{i}{\Gamma^a}{\Gamma^{\mu\nu}}{\lambda^{j}}
{\chi_{j}})\nabla_a Y\cr
= & {2 \over \alpha^{\prime 3}}
{\cal O}^{\prime}_{ij}{\bar \lambda^i}\Gamma^a
{\lambda^j}
\nabla_aY & (what)\cr}
$$
\noindent where
$$
{\cal O}^{\prime}_{ij}({T}) {\equiv} {{1}\over{4}}\,
{\int}{d^4}{x}{\sqrt{g}}\,e^{-2\phi}{\chi^\dagger_{i}}
{\Gamma^{\mu\nu}}{T_{\mu\nu}}{\chi_{j}}. \eqno(mapone)
$$
The lambda bilinear appearing in (\call{what}) can
be seen, using
(\call{vacuum}), to be equivalent to
$\psi^i\psi^j$ when acting on a vacuum state.
Substituting the formula (\call{fmode}) for ${\chi_{i}}$
one has, after some algebra
$$
{\cal O}^{\prime}_{ij}({T}) = {\int_{X}}\,{\rm
tr}({\delta_i}{A}{\wedge}{\delta_j}
{A}){\wedge}{T_+},\eqno(aat)
$$
where $T_+$ is the self-dual part of $T$.
It is
easily checked that ${\cal O}^{\prime}_{ij}$
is not closed and so
does not represent a cohomology class on ${\cal M}_{N}$.

This is remedied by the observation that (\call{hxx}) is not the only term
in ${S_{10}}$ which gives rise to the coupling of a spacetime axion to
worldbrane fermions. Because of the bilinear term in the expansion
(\call{atan}) for ${A_a}$, such couplings
also arise from the ten-dimensional term
$$
{2 \over \alpha^{\prime 3}}e^{-2
\phi}{\partial_M}{B_{NP}}{\omega^{MNP}_{3Y}}.\eqno(bomega)
$$
\noindent From the expansion for ${A_a}$, the relevant term in $\omega_{3Y}$
is
$$
{\omega^{3Y}_{a\mu\nu}} = -{1 \over 2} {\rm tr}\,({\phi_{ij}}{\bar\lambda}^{i}
{\Gamma_a}{\lambda^j}{F_{\mu\nu}}).\eqno(phif)
$$
\noindent This formula may then be used to reduce (\call{bomega})
to a coupling in ${S_{\rm eff}}$. The result may be added to
(\call{what}) to give the total coupling of a single spacetime axion of the
form (\call{axion}) to two worldbrane fermions:
$$
{\cal L}_{\rm int}=
{2 \over \alpha^{\prime 3}}
{\cal O}_{ij}{\bar \lambda^i}\Gamma^a
{\lambda^j}
\nabla_aY
\eqno(coup)
$$
\noindent where
$$
{\cal O}_{ij}({T}) = {\int_{X}}\,{\rm tr}({\delta_i}{A}{\wedge}
{\delta_j}{A} -{\phi_{ij}}{F}){\wedge}{T_+}\eqno(mfinal)
$$
(\call{mfinal}) has several important properties. The first
is that
${\cal O}$ is closed on ${\cal M}_{N}$:
$$
{\partial_{[i}}{\cal O}_{jk]}=  -{\int_{X}}\,{\rm tr} ({D}{\phi_{[ij}}{\wedge}
{\delta_{k]}}{A} + {\phi_{[ij}}{s_{k]}}{F}){\wedge}{T_+} = 0 \eqno(dm)
$$
\noindent upon integration by parts on ${X}$. Secondly, if
${T_+}$ is trivial in ${H^2}({X})$ so that ${T_+} = {dU}$
one has
$$\eqalignno{
{\cal O}_{ij}({dU})= & {\int_X}\,{\rm tr}
({D}{\delta_{[i}}{A}{\wedge}{\delta_{j]}}{A} +
{s_{[i}}{\delta_{j]}}{A}{\wedge}{F}){\wedge}{U}\cr
= & -2{\partial_{[i}}{\int_X}\,{\rm
tr}\,({\delta_{j]}}{A}{\wedge}{F}){\wedge}{U}
 & (tdu)\cr}
$$
\noindent i.e. the image of an exact form on ${X}$  is an exact
form on ${\cal M}_{N}$. Thus (\call{mfinal}) gives a
map from the cohomology of ${X}$ into the cohomology of ${\cal M}_{N}$.
Using Poincare duality (\call{mfinal}) may be written:
$$
{\cal O}_{ij}({\Sigma}) = {\int_{\Sigma}}\,{\rm tr}\,({\delta_i}{A}{\wedge}
{\delta_j}{A} - {\phi_{ij}}{F})\eqno(pdual)
$$
\noindent where ${\Sigma}$ is the surface Poincare dual to ${T_+}$. This is
a standard expression [\cite{donref}] for the Donaldson map from
${H_2}({X}){\rightarrow}{H^2}({\cal M}_{N}({X}))$ in terms of
differential forms, and is identical to that derived in the context
of topological quantum field theory by Witten [\cite{WIT.2}].

\item{\bf B.}{\bf  Derivation from K\"ahler Geometry.}

In this subsection
we will provide an alternate derivation of (\call{coup}) which is less direct,
but shorter and provides some geometrical insight. For these purposes
it is convenient to view the solution (\call{sdual})--(\call{metax}) not
as ${N}$ fivebranes on ${X}$, but as a ``compactification'' from ten to six
dimensions. The  low-energy action then contains, in addition to ${Z^i}$,
complex massless moduli fields ${Y^{\alpha}}$ that parameterize
the complexified K\"ahler cone ( a subset of  ${H}^2(X,C))$.
The imaginary part of ${Y^{\alpha}}$
is the axion associated to the harmonic two form ${T_{\alpha}}$ on
${X}$ (The $\alpha$ index was suppressed in the previous subsection.).
Six-dimensional supersymmetry then implies that the metric appearing
in the kinetic term for the moduli fields is K\"ahler, or equivalently
in complex coordinates
$$
{J}_{{I}{\bar J}} =i {\partial}_{I}{\partial}_{{\bar J}}{\cal K}.\eqno(jform)
$$

To give an expression for ${\cal K}$, we note that on a K\"ahler manifold a
closed $(p,p)$ form $H_{p,p}$ is locally the curl of a $2p-1$ form:
$$
H_{p,p} = d G_{2p-1} = (\partial + \bar \partial)(G_{p,p-1} + G_{p-1,p}).
\eqno(gcurl)
$$
Since the left-hand side of (\call{gcurl}) is of type $(p,p)$,
$$
\partial G_{p,p-1} = \bar \partial G_{p-1,p} = 0, \eqno(igiveup)
$$
so that locally $G_{p,p-1} = \partial F_{p-1,p-1}$.
We conclude that locally a closed $(p,p)$ form can always be written in the
form
$$
H_{p,p} = i \partial \bar \partial F_{p-1,p-1} . \eqno(ddbar)
$$
$F$ is real if $H$ is, and is determined up to a closed $(p-1,p-1)$ form.

${\cal K}$ is then given by
$$
{\cal K}={1 \over 2}\int_X E \wedge J-{\rm ln }\int_X J\wedge J
\eqno(whytwo)
$$
where $J$ is the K\"ahler form on $X$ and $E$ is a solution of
$$
{\rm tr}F \wedge F=i\partial \bar \partial E.
\eqno(ffetwo)
$$
$E$ is related to the two-dimensional WZW action and can not be simply
expressed as a function
of $A$. A formula for ${\cal K}$ as a conformal field theory correlation
function is given in [\cite{perstrom}].

The second variation of
${\cal K}$ can be computed by noting that
$$
{\partial}_{I}{\partial}_{{\bar J}}{\rm tr}F\wedge F=2
\bar \partial \partial {\rm tr} ({\delta_I}
{A}{\wedge}{\delta}_{\bar J}{A} - {\phi}_{{I}{\bar J}}{F}). \eqno(ddbarff)
$$
This determines the second variation of $E$ up to a closed two-form on $X$
times
a closed two-form on ${\cal M}_N$. The ambiguity in the definition of $E$ may
thus be fixed
so that
$$\eqalignno{
%{\partial_I}{\cal K}={\int_X}\,{\rm tr}({F}{\wedge}{\delta_I}{A})
%{\wedge}{\bar \partial K}\cr
i{\partial}_{I}{\partial}_{{\bar J}}{\cal K}= & - {\int_X}\,{\rm tr}({\delta_I}
{A}{\wedge}{\delta}_{\bar J}{A} - {\phi}_{{I}{\bar J}}{F}){\wedge}J
\cr
= & {\int}{\sqrt{g}}\,{J^{\mu\nu}}\,{\rm tr}\,{\delta_I}{A_\mu}
{\delta}_{\bar J}{A_\nu}\cr
= & {\cal J}_{{I}{\bar J}} &(dk)\cr}
$$
\noindent where in the last line we have used ${J}{\cdot}{F} = {0}$.
The coupling of ${Y}$ to two ${\lambda^\prime}$'s is then determined
by supersymmetry to be proportional to
the mixed Christoffel connection (as in  (\call{swb}) with an index
lowered) on
${H}^2(X,C)\times {\cal M}_N$:
$$
{\cal L}_{\rm int} = {-2 i \over \alpha^{\prime 3}}
{\nabla_a}{Y^{\alpha}}{\bar{\lambda^\prime}}^{\bar J}{\Gamma^a}
{\lambda^\prime}^{I}\Gamma_{{\bar J}I \alpha}\eqno(lint)
$$
In K\"ahler geometry, the Christoffel connection is given by
$$
\Gamma_{{\bar J}I \alpha} = {\cal K},_{{\bar J}I \alpha}.\eqno(ok)
$$
Differentiating (\call{dk}) one more time and
using ${\partial_{\alpha}}{J} = {T_{\alpha}}$ we easily recover
the formula (\call{mfinal}) of the previous section
$$
-i \Gamma_{{\bar J}I \alpha} = {\int_X}\,{\rm tr}\,({\delta_I}
{A}{\wedge}{\delta}_{\bar J}{A} - {\phi_{{I}{\bar J}}} {F})
{\wedge}{T_{\alpha}}={\cal O}_{{\bar J} I\alpha } \eqno(recov)
$$
except for the absence of a projection on to the self-dual part
of $T_\alpha$. This difference can be accounted for if $\lambda^\prime$
is related to $\lambda$ of the previous section by the field redefinition
$$
\lambda^\prime = e^{- i X_\alpha Y^\alpha} \lambda \eqno(fielddef)
$$
where $X_\alpha \equiv \int T_\alpha \wedge J / \int J \wedge J$.

\item{\bf C.}{\bf The Donaldson Map as the Second Chern Class}

It is known [\cite{donref}, see also \cite{kan,basi}]
that ${\cal O}$ can be written as integrals of
${\rm tr}{\cal F}^2$ for a certain curvature $\cal F$. The fact that
$\cal O$ couples to axions then strongly suggests that the observations
in this paper are connected with the structure of
anomalies in string theory. While
we do not understand this connection, in the hope that it
might be understood later we record here this representation of $\cal O$.
Introduce a connection ${\cal D}$ (on the universal bundle over ${\cal
M}_N$) by
$$
{\cal D} = {d}{Z^i}{s_i} + {dx^{\mu}}{D}_{\mu}. \eqno(bigd)
$$
\noindent Then the associated curvature ${\cal F} = {\cal D}^2$ has
components
$$\eqalignno{
{\cal F}_{ij}=& {\phi_{ij}}\cr
{\cal F}_{i\mu}=&{\delta_i}{A}_{\mu}\cr
{\cal F}_{\mu\nu}=& F_{\mu\nu}. &(bigf)\cr}
$$
\noindent Now consider the integral ${c_2}({\Sigma})$ of the second Chern
class of ${\cal F}$ over a four surface ${\Sigma}$ in
${\cal M}_{N}\times{X}$
$$
{c_2}({\Sigma}) = {1 \over 8\pi^2}{\int_{\Sigma}}\,{\rm tr}\,{\cal
F}{\wedge}{\cal F}
\eqno(chern)
$$
\noindent If ${\Sigma}$ is a product of a two surface ${\Sigma_{\cal M}}$ in
${\cal M}_{N}$ with a two surface ${\Sigma_X}$ in ${X}$ one finds
$$
{c_2}({\Sigma_{\cal M}}{\times}{\Sigma_X}) =
-{1 \over 8 \pi^2}{\int}_{\Sigma_{\cal M}}\,
{\cal O}_{ij}
({\Sigma_X}){d}{Z^i}{\wedge}{d}{Z^j}.\eqno(cm)
$$
\noindent i.e. the Donaldson map ${H_2}({X})\rightarrow {H^2}({\cal M}_{N})$ is
a period of the second Chern class. A similar result holds for the
maps ${H}_{\alpha}({X}){\rightarrow}{H}^{4- \alpha} ({\cal M}_{N}).$

\item{\bf D.}{\bf  The Donaldson Polynomial}

A physical realization of the Donaldson polynomial may be obtained by
considering multiple axion scattering. Let $|m \rangle $ be the ground state
corresponding to the top rank form on ${\cal M}_{N}$:
$$
{|m} \rangle  = {\epsilon}_{i_1}{\cdots}{}_{i_m}\,{\lambda}^{i_1}{\cdots}
{\lambda}^{i_m}|0 \rangle .\eqno(top)
$$
\noindent The amplitude
for scattering ${p}$ axions associated to
the classes
${T_1}{\cdots}{T_p}$ off
the state $|0 \rangle $ and winding up in the state $|m \rangle $ is
proportional to
$$
{A}({T_1},{\cdots}{T_p}) = {\langle m|}{\cal O}^{T_1}
                      {\cdots}{\cal O}^{T_p}|0 \rangle .\eqno(multi)
$$
\noindent It is easily seen that this reduces to
$$
{A}({T_1},{\cdots},{T_p}) = {\int_{{\cal M}_{N}}}{\cal O}({T_1}){\wedge}
{\cdots}{\wedge}{\cal O}({T_p})\eqno(poly)
$$
\noindent which is the Donaldson polynomial.

While our derivation from string theory of (\call{mfinal,poly}) was only valid
for ${c_1}(X) \geq 0$,
it is known [\cite{donref}] that (\call{mfinal}) and (\call{poly}) are
representations of the Donaldson map and polynomial for any algebraic
${X}$. It would be interesting to try to extend our derivation to the
more general case.

\head{\bf IV.CONCLUSION}

We have shown that the Donaldson map appears explicitly as a coupling
in the low-energy action for heterotic string theory in the soliton
sector. This implies the Donaldson polynomial can be measured by
scattering massless fields and solitons. This realization leads to
concrete formulae for the Donaldson map and polynomial which are
equivalent to, and provide a new perspective on, formulae derived
by Witten in the framework of topological Yang-Mills theory. It also led to
an interpretation of
the Donaldson map as a K\"ahler-Christoffel
connection on ${H}^2(X,C)\times {\cal M}_N(X)$.

The fact that this scattering is a map between cohomology classes
was insured by zero-momentum worldbrane supersymmetry, which acts
like the exterior derivative on ${\cal M}_{N}$.
%closed, the worldbrane part of the operator in (\call{coup}) commutes with
%the supercharge (as well as the Hamiltonian) and therefore can only
%induce transitions among the ${N}$-fivebrane groundstates.
This should
be contrasted with topological
Yang-Mills theory where the exterior derivative on the instanton moduli
space is constructed in terms of a BRST operator.

Our work suggests a number of generalizations and applications. Perhaps
this connection can provide new insights into, or stringy
interpretations of, the various theorems on the structure of
four-manifolds which follow from Donaldson's work.
Alternately, the remarkable properties of
the Donaldson polynomial may translate into interesting properties
of the fivebrane-axion S-matrix, or even have implications for the closely
related problem of instanton-induced supersymmetry breaking in string theory.

\head{\bf ACKNOWLEDGMENTS}

This work was supported in part by DOE Grant DE-AT03-76ER70023 and
NSF Grant PHY90-00386.
We are grateful to P. Bowcock,  R. Gregory,  D. Freed, S. Giddings,
D. Moore, C. Vafa and S. T. Yau for useful discussions.

\references

\refis {STR.26} A. Strominger, {\it Superstrings with Torsion}, Nuclear
Physics {\bf B274}, 253 (1986).

\refis {basi} L. Balieu and I. M. Singer, Nucl. Phys. N (Proc. Suppl.)
{\bf 5B}, 12 (1988).

\refis {STR.55} A. Strominger, {\it Heterotic Solitons}, Nuclear Phys. B,
{\bf 343}, 167 (1990).

\refis {strb} C. Callan, J. Harvey and A. Strominger, {\it Worldsheet
Approach to Heterotic Instantons and Solitons}, Nucl. Phys. B359 (1991) 611.

\refis {wbrane} C. Callan, J. Harvey and A. Strominger, {\it Worldbrane Actions
for String Solitons}, PUPT-1244, EFI-91-12, (1991).

\refis {WIT.2} E. Witten, {\it Topological Quantum Field Theory}, Comm.
Math Phys. {\bf 117}, 353 (1988).

\refis {wittena} E. Witten, {\it Constraints on Supersymmetry Breaking},
Nucl. Phys. B {\bf 302}, 253 (1982).

\refis {sjrey} S. J. Rey,
%{\it Axionic String Instantons and Their
%Low-Energy Implications}, UCSB preprint UCSB-TH-89-23, (1989) and
{\it Confining Phase of Superstrings and Axionic Strings}, Phys. Rev.
D {\bf 43} (1991) 439.

\refis {teinep} R. Nepomechie, Phys. Rev. D {\bf 31} (1985)
1921; C. Teitelboim, Phys. Lett. B {\bf 176} (1986) 69.

\refis {kan} H. Kanno, Z. Phys. C {\bf 43} (1989) 477.

\refis {zum} B. Zumino, Phys. Lett. B {\bf 69} (1977) 369.

\refis {HP.1} J. Hughes and J. Polchinski, Nucl. Phys. B.

\refis {HLP.2} J. Hughes, J. Liu and J. Polchinski, {\sl Phys. Lett.}
{\bf 180B} (1986), 370.

\refis {HST.1} P. S. Howe, G. Sierra and P. K. Townsend, Nucl. Phys.
{\bf B221} (1983) 331.

\refis {SITO.1} G. Sierra and P. K. Townsend, {\it The Gauge-Invariant}
${N=2}$ {\it Supersymmetric} ${\sigma}$
{\it Model with General Scalar Potential}
Nucl. Phys. B {\bf 233}, 289 (1984).

\refis {yau} S. T. Yau, private communication.

\refis {oova} A related expression
has been previously considered for different purposes by H. Ooguri and C.
Vafa (private
communication).

\refis {donref} S. K. Donaldson and P. B. Kronheimer,
{\it The Geometry of Four-Manifolds},
Oxford University Press (1990) and references therein.

\refis {perstrom} V. Periwal and A. Strominger, Phys. Lett. B235 (1990) 261.

\endreferences
\endit
\end